\documentclass[prl, reprint,showpacs,  superscriptaddress]{revtex4-1}

\usepackage{color,xcolor}

\usepackage{footnote}
\usepackage[colorinlistoftodos,bordercolor=red!50,backgroundcolor=red!10,linecolor=red!50,textsize=tiny]{todonotes}
\usepackage{changes}

\usepackage[utf8]{inputenc}
\usepackage[T1]{fontenc}
\usepackage[english]{babel}  
\usepackage{times}
\usepackage{bm}
\usepackage{ulem}

\usepackage[colorlinks=true,linkcolor=blue,citecolor=blue,urlcolor=blue,pdfauthor={ },pdftitle={ },pdfsubject={ },pdfkeywords={ }]{hyperref}
\usepackage{amssymb,amsmath,amsfonts,amsthm}
\usepackage{mathtools}
\usepackage{verbatim}
\usepackage{enumerate}
\usepackage{graphicx}
\usepackage{hyperref}
\usepackage{bbm}
\usepackage{booktabs}

\usepackage[caption=false]{subfig}

\usepackage{color}
\definecolor{dred}{rgb}{.8,0.2,.2}
\definecolor{ddred}{rgb}{.8,0.5,.5}
\definecolor{dblue}{rgb}{.2,0.2,.8}
\definecolor{dgreen}{rgb}{.2,0.5,.2}

\usepackage{multirow}

\def\be{\begin{eqnarray}}
\def\ee{\end{eqnarray}}

\usepackage{times}

\definecolor{Pr}{rgb}{0.4,0.3,0.9}

\begin{document}
\title{QCSH: a Full Quantum Computer Nuclear Shell-Model Package}

\author{Peng Lv}
\thanks{These authors contributed equally to this work.}
\affiliation{State Key Laboratory of Low-Dimensional Quantum Physics and Department of Physics, Tsinghua University, Beijing 100084, China}

\author{Shi-Jie Wei}
\thanks{These authors contributed equally to this work.}
\affiliation{Beijing Academy of Quantum Information Sciences, Beijing 100193, P. R. China}

\author{Hao-Nan Xie}
\affiliation{State Key Laboratory of Low-Dimensional Quantum Physics and Department of Physics, Tsinghua University, Beijing 100084, China}
\affiliation{Beijing Academy of Quantum Information Sciences, Beijing 100193, P. R. China}

\author{Gui-Lu Long}
\email{gllong@tsinghua.edu.cn}
\affiliation{State Key Laboratory of Low-Dimensional Quantum Physics and Department of Physics, Tsinghua University, Beijing 100084, China}
\affiliation{Beijing Academy of Quantum Information Sciences, Beijing 100193, P. R. China}

\begin{abstract}
Nucleus is a typical many-body quantum system. Full calculation of a nuclear system in a classical computer is far beyond the capacity of current classical computers. With fast development of hardware, the prospect of using quantum computers in nuclear physics is closing. Here, we report a full quantum package, QCSH,  for solving nuclear shell-model in a quantum computer. QCSH uses the linear combination of unitaries formalism of quantum computing, and performs all calculations in a quantum computer. The complexities of qubit resource, number of basic gates of QCSH, are both polynomial to the nuclear size. QCSH can already provide meaningful results in the near term. 
As examples, the binding energies of twelve light nuclei,  $^{2}$H, $^{3}$H, $^{3}$He, $^{4}$He, $^{6}$Li, $^{7}$Li, $^{12}$C, $^{14}$N, $^{16}$O, $^{17}$O,  $^{23}$Na and $^{40}$Ca are calculated using QCSH in a classical quantum emulator.  The binding energy of Deuteron has already been experimentally studied using QCSH on a  superconducting quantum computing device. QCSH not only works in near-term quantum devices, but also in future large-scale quantum computers. With the development of quantum devices, nuclear system constitutes another promising area for demonstrating practical quantum advantage.
\end{abstract}

\maketitle

\section{Introduction}
Nucleus,  an important layer of material, is a quantum many-body system  composed of protons and neutrons. The calculation in nuclear physics is tremendously difficult due to the huge state space involved. The study of nuclear system is even more difficult than atomic and molecular quantum many-body system because the complexities of the configuration space and uncertainties of the nuclear interaction  \cite{jastrow1951nucleon,heyde1994nuclear}. Nuclear forces are treated as effective interaction potentials in realistic models to study  the properties of nuclei such as energy levels, electromagnetic transition rates and reaction cross sections. The nuclear shell model is a typical example of this kind of realistic models. In nuclear shell model,  each nucleon moves independently in an effective average field, and the nucleons interact with each other via residual interaction. Shell-model explains the formation of magic numbers and  nuclear collective motion with great success \cite{mayer1950nuclear, haxel1949magic,elliott1958collective}. Advanced shell-model codes,  such as the NuShellX \cite{brown2014shell} are extensively used for various calculations.

Full calculation of nuclear shell-model, with all nucleons and all shells, in a classical computer is intractable, because of the limitations of computational resources. For instance, NuShellX code can treat a matrix with a dimension up to the order of  1 $\times 10^8$. Currently, nuclear shell-model calculations are performed using several restrictions, for instance, only valence shells are considered and the inner shells are treated as frozen. Even with such restrictions, shell-model  can only calculate the structure of light nuclei. For medium and heavy nuclei, calculations using valence nucleons are still not possible.  The combined uncertainty in the nuclear interaction and the approximations in the calculation, makes the study of nuclear system extremely challenging.

In another front, fast development in quantum computing is attracting worldwide attention \cite{arute2019quantum}. For simulating a quantum many-body system with size $N$, quantum computer  only consumes $O(poly(N))$ resources, while a classical computer has to use $O(2^N)$ resources.  With the huge progress in quantum computer hardware, rudimentary quantum computers with  100 qubits are now available. Though quantum  hardware are still far from the sophisticated fault-tolerant quantum computer, they can already complete useful tasks \cite{preskill2018quantum,kandala2017hardware,google2020hartree,harrigan2021quantum}.

The prospect of using quantum computer to study quantum system is more than ever close. In  chemistry field,  the variational quantum eigensolver (VQE) \cite{peruzzo2014variational} and the full quantum eigensolver (FQE)\cite{wei2020full} approach have been proposed to find the eigenvalues and eigenstates of a molecule.
In the quantum-classic hybrid model, VQE,  the computation process is switched iteratively between a classical computer and a quantum computer. Recently in nuclear physics, VQE has been applied to claculate the binding energy of the deuteron nucleus \cite{dumitrescu2018cloud, lu2019simulations}.

In contrast, FQE  is  full quantum computer algorithm \cite{wei2020full}, which is based on the linear combination of unitaries formalism \cite{gui2006general}.  FQE performs all calculations on a quantum computer, avoiding the slow and tedious switching between classic and quantum computers. FQE has been demonstrated in chemistry calculation \cite{wei2020full}, and attracted attention in recent years \cite{tranter2019ordering, chivilikhin2020mog, robert2021resource, smart2021quantum, jin2020query, hu2021recent}.

In this paper, we present a full quantum computer package, QCSH,  for solving the nuclear shell model in a quantum computer. In the following, we firstly present the framework of the quantum computer nuclear shell-model. Then, we verify QCSH by showing that it gives identical results as those from the exact diagonalization in a classical computer through examples. To estimate the influence of imperfection of near-term quantum computers, we also perform QCSH calculation with quantum noises. 
We demonstrate the use of QCSH by applying it to find the binding energies of twelve light nuclei, $^{2}$H, $^{3}$H, $^{3}$He, $^{4}$He, $^{6}$Li, $^{7}$Li, $^{12}$C, $^{14}$N, $^{16}$O, $^{17}$O, $^{23}$Na and $^{40}$Ca. Moreover, we experimentally demonstrate the algorithm on a superconducting  quantum hardware  to find the binding energy of a deuteron. Finnaly, the quantum resources consumed in this algorithm is analyzed which indicates a quantum speedup.

\section{The QCSH framework}
\subsection{Hamiltonian and model}
	
 We use a nuclear shell-model with an effective average field with spin-orbit coupling \cite{mayer1950nuclear,heyde1994nuclear}. In addition to the average field, the nucleons are also interacting through the residual interaction. 	For a given nucleus $^{A}_{Z}X$, the Hamiltonian can be written in first quantization formalism as
	\begin{equation}
	\begin{aligned}
		H = &\sum_{i=1}^{Z}\frac{\vec{p}_{i}^{2}}{2M_{p}} + \sum_{j=Z+1}^{A}\frac{\vec{p}_{j}^{2}}{2M_{n}} + \sum_{k=1}^{A}V(\vec{r_{k}}) \; + \\ 
		&\sum_{1\leq i_{1} < i_{2} \leq Z} \frac{e^{2}}{4\pi \epsilon_{0}}\frac{1}{|\vec{r}_{i_{1}}-\vec{r}_{i_{2}}|} + \frac{1}{2}\sum_{k_{1} \neq k_{2}}u(k_{1}, k_{2}),
	\end{aligned}
	\label{hamiltonian}
	\end{equation}
where $M_{p}=938.3{\rm \;MeV}/c^{2}$, $M_{n}=939.6{\rm \;MeV}/c^{2}$ are the mass of proton and neutron respectively. ${ V}(\vec{ r}_{ k})$ is the average field, and $u({ k}_{1}, { k}_{2})$ is the residual two-body interaction. The first two terms are kinetic energy of protons and neutrons, respectively. The fourth term  is the Coulomb potential between protons. To transform the Hamiltonian from first quantization into second quantization, we choose the single particle orbits for protons and neutrons $\{\psi_{\pi, i}\}_{i=1}^{N_{\pi}}$ and $\{\psi_{\nu, j}\}_{j=1}^{N_{\nu}}$ \cite{fetter2012quantum} respectively, where $N_{\pi}$ and $N_{\nu}$ are the number of single particle orbits taken in the calculations.

Once the single-particle orbits have been selected, the Hamiltonian can be written in the second quantization form as
\begin{equation}
\begin{aligned}
	H=&\sum_{\alpha '}\sum_{\alpha}g_{\pi}(\alpha ', \alpha)a_{\pi,\alpha '}^{\dagger}a_{\pi,\alpha}\otimes I_{\nu} \;+\\ 
	& \sum_{\alpha_{1}' \alpha_{2}'}\sum_{\alpha_{1} \alpha_{2}} h_{\pi}(\alpha_{1}', \alpha_{2}',\alpha_{1},\alpha_{2})a_{\pi,\alpha_{1}'}^{\dagger}a_{\pi,\alpha_{2}'}^{\dagger}a_{\pi,\alpha_{2}}a_{\pi,\alpha_{1}}\otimes I_{\nu}\;+\\
	&\sum_{\beta '}\sum_{\beta}g_{\nu}(\beta', \beta)I_{\pi}\otimes a_{\nu,\beta'}^{\dagger}a_{\nu,\beta}\;+ \\ 
	&\sum_{\beta_{1}' \beta_{2}'}\sum_{\beta_{1} \beta_{2}} h_{\nu}(\beta_{1}', \beta_{2}',\beta_{1},\beta_{2})I_{\pi}\otimes a_{\nu,\beta_{1}'}^{\dagger}a_{\nu,\beta_{2}'}^{\dagger}a_{\nu,\beta_{2}}a_{\nu,\beta_{1}} \;+\\
&\sum_{\alpha ' \beta '}\sum_{\alpha \beta}h_{\pi\nu}(\alpha ', \alpha, \beta ', \beta)a_{\pi,\alpha '}^{\dagger}a_{\pi,\alpha}\otimes a_{\nu,\beta '}^{\dagger}a_{\nu,\beta},
\end{aligned}
\label{hsq}
\end{equation}
where $\alpha$, $\alpha '$, $\alpha_{1}$, $\alpha_{1}'$, $\alpha_{2}$, $\alpha_{2}'$ $\in$ $\{1, 2, ..., N_{\pi}\}$, $\beta$, $\beta '$, $\beta_{1}$, $\beta_{1}'$, $\beta_{2}$, $\beta_{2}'$ $\in \{1, 2, ..., N_{\nu}\}$. $a^{\dagger}$, $a$ are the creation and annihilation operators acting on a single particle state respectively, and the operators of protons and neutrons are denoted  by subscript $\pi$ and $\nu$ respectively. The one-body and two-body interactions in the hamiltonian can be calculated by
\begin{equation}
\begin{aligned}
	&g_{\pi}(\alpha ', \alpha)=\langle \psi_{\pi, \alpha'}|\frac{\vec{p}^{2}}{2M_{p}}+V(\vec{r})|\psi_{\pi, \alpha} \rangle, \\
	&g_{\nu}(\beta ', \beta)=\langle \psi_{\nu, \beta'}|\frac{\vec{p}^{2}}{2M_{n}}+V(\vec{r})|\psi_{\nu, \beta} \rangle, \\
	&
	\begin{aligned}
		h_{\pi}(\alpha_{1}', \alpha_{2}',\alpha_{1},\alpha_{2})=&\langle \psi_{\pi, \alpha_{1}'}|\langle \psi_{\pi, \alpha_{2}'}|\frac{1}{2}\Big\{\frac{e^{2}}{4\pi \epsilon_{0}}\frac{1}{|\vec{r}_{\pi_{1}}-\vec{r}_{\pi_{2}}|} \\
	&+u(\pi_{1}, \pi_{2})\Big\}|\psi_{\pi, \alpha_{1}}\rangle |\psi_{\pi,\alpha_{2}}\rangle,
	\end{aligned} \\
	&h_{\nu}(\beta_{1}', \beta_{2}',\beta_{1},\beta_{2}) = \langle \psi_{\nu, \beta_{1}'}|\langle \psi_{\nu, \beta_{2}'}|\frac{1}{2}u(\nu_{1}, \nu_{2})|\psi_{\nu, \beta_{1}}\rangle |\psi_{\nu, \beta_{2}}\rangle, \\
	&h_{\pi\nu}(\alpha ', \alpha, \beta ', \beta) = \langle \psi_{\pi, \alpha'}|\langle \psi_{\nu, \beta'}|u(\pi, \nu)|\psi_{\pi, \alpha}\rangle |\psi_{\nu, \beta}\rangle.
\end{aligned}
\label{hsqc}
\end{equation}

Here we choose the average field $V(\vec{r})$ as the Woods-Saxon potential with a spin-orbital coupling term \cite{bohr1969single} 
\begin{equation}
\begin{aligned}
	&V(\vec{r}) = U_{c}(r) - 2\lambda \; \Big(\frac{\hbar}{2Mc}\Big)^{2} \cdot \frac{dU_{c}(r)}{dr}\vec{s}\cdot \vec{l} \; , \\
	&U_{c}(r) = U_{0}(1\pm \kappa \frac{N - Z}{N+Z})/(1+\exp(\frac{r-r_{0}A^{1/3}}{a_{0}})),
\end{aligned}
\label{woodsaxon}
\end{equation}
where $\kappa = 0.67$, $r_{0}=1.27$ fm, $a_{0}=0.67$fm, $\lambda = 32$ fm$^{-1}$ and $M = (M_{p}+M_{n})/2=938.95$ M $eV/c^{2}$. The average field of protons and neutrons are not exactly the same, as for the $\pm \kappa$ of Woods-Saxon potential in (\ref{woodsaxon}), protons take the positive sign and neutrons take the negative sign. $\vec{s}$ and $\vec{l}$ are the single particle spin and orbital angular momentum operators respectively.  $\vec{s}\cdot \vec{l} = l/2$ for $j=l+1/2$ and $ -(l+1)/2$ for $j=l-1/2$. The average field can be calculated also from relativistic mean field approach\cite{meng2021relativistic}.

A major component of the two-body interaction $u(k_{1}, k_{2})$, which can be  written as \cite{heyde1994nuclear}
\begin{equation}
\small
\begin{aligned}
	H_{\rm pair} =& \frac{-G}{4} \Big[ \sum_{j, m_{j}, m_{j} '}(-1)^{2j+m_{j}+m_{j} '}a^{\dagger}_{\pi, j,m_{j}}a^{\dagger}_{\pi, j,-m_{j}}a_{\pi, j,m_{j} '}a_{\pi, j,-m_{j} '} \otimes I_{\nu} \\
	&+\sum_{j, m_{j}, m_{j} '}(-1)^{2j+m_{j}+m_{j} '}I_{\pi}\otimes a^{\dagger}_{\nu, j,m_{j}}a^{\dagger}_{\nu, j,-m_{j}}a_{\nu, j,m_{j} '}a_{\nu, j,-m_{j} '} \Big] \; ,
\end{aligned}\label{pairing}
\end{equation}
where $G=0.25$MeV. Thus, in nuclear shell-model, one can first give a Hamiltonian in the form of Eq.(\ref{hsq}), taking an average potential to calculate the single particle orbits, and adopt a form of residual interactions from either a microscopic model or from a phenomenological model. Then calculate the matrix elements by  Eq. (\ref{hsqc}). In the examples of this work, we take the interaction as given in Eq. (\ref{hsq}).

\subsection{Encoding the Quantum state into Qubits}

Fermions obey the Pauli principle, and hencefore one qubit is enough to represent the possession  (1) or absence (0) of a nucleon in a single particle state. The state of a nucleus with $Z$ protons and $N$ neutrons could be written as the superposition of the basis states, 
\begin{equation}
\begin{matrix} N_{\pi}\; qubits \\ \overbrace{\begin{matrix} \underbrace{|0100...101\rangle} \\ Z\; {\rm protons} \end{matrix}} \end{matrix} \begin{matrix} N_{\nu}\; qubits \\ \overbrace{\begin{matrix} \underbrace{|1001...110\rangle} \\ N\; {\rm neutrons} \end{matrix}} \end{matrix}\;\;.
\label{qubitrep}
\end{equation}
where $N_{\pi}$	are proton single particle orbits, and $N_{\nu}$ are neutron single particle orbits.
After encoding quantum states on the qubits, we use the Jordan-Wigner(J-W) transformation \cite{jordan1993paulische} to map the creation and annihilation operators into the Pauli operators as
\begin{equation}
\begin{aligned}
	a_{k}^{\dagger} &= c_{k}|1\rangle_{k}\langle 0|  \\
	&= \frac{1}{2}Z_{1}\otimes Z_{2}\otimes...Z_{k-1}\otimes (X_{k}-iY_{k}) \otimes I_{k+1} \otimes ... \\
	a_{k} &= c_{k}|0\rangle_{k}\langle 1|  \\
	&= \frac{1}{2}Z_{1}\otimes Z_{2}\otimes...Z_{k-1}\otimes (X_{k}+iY_{k}) \otimes I_{k+1} \otimes ...
\end{aligned}
\end{equation}
where subscript $k$ denotes the single particle state labeling, $c_{k} = (-1)^{n_{1}+n_{2}+...+n_{k-1}} \in \{ -1, +1\}$ and $n_{1}, n_{2}, ...n_{k-1} \in \{0, 1\}$ are the particle number in each single particle state. $X$, $Y$, $Z$ are the Pauli gates \cite{nielsen2000quantum}. Nucleus contains both protons and neutrons and the transformation is carried for protons and neutrons separately. The operators should  also be labeled with the kind of nucleon such as $a^{\dagger}_{\pi, k}$, $a_{\pi, k} $, $a^{\dagger}_{\nu, k}$ , $a_{\nu, k} $, $I_{\pi}$ and $I_{\nu}$.

After the J-W transformation, the Hamiltonian in the second quantization formalism is represented as
\begin{eqnarray}
	H = \sum_{k=0}^{M-1} \alpha_{k} H^{g}_{k},\label{elcu}
\end{eqnarray}
where $H_{k}^{g}$ is unitary operator composed of  Pauli operators. Eq. (\ref{elcu}) can be expressed in terms of linear combination of unitary operators as given in Refs \cite{long2011duality}.

\subsection{Quantum Gradient Descent Iteration}
Instead of diagonalizing the Hamiltonian directly, we use the gradient descent iteration as in the FQE approach \cite{wei2020full} using the LCU formalism \cite{gui2006general} in  QCSH. We start from a trial wavefunction $|\phi^{(t)}\rangle$, and  the average energy of the system is $\langle E\rangle = \langle \phi^{(t)}|H|\phi^{(t)}\rangle$. Taking the  gradient of the average energy, we have $\nabla \langle \phi^{(t)}|H|\phi^{(t)}\rangle = 2 H |\phi^{(t)}\rangle$. The next step will be iterated as
\begin{equation}
	|\phi^{(t+1)}\rangle \propto (I-2\gamma H)|\phi^{(t)}\rangle,
\label{qdescent}
\end{equation}
where $\gamma$ is a scaling factor characterising the amount of the step  in the iteration, usually $\gamma > 0$. The determination of  the value of $\gamma$ will be briefly discussed in  Appendix A.

Denote  the iteration operator $T = I-2\gamma H$. In general, $T$ is not unitary, and it can be implemented using the LCU formalism \cite{long2011duality}. Explicitly,  we add an ancillary register with  $m=\lceil \log_{2}M \rceil$ qubits in  state $|\psi_s\rangle=\frac{1}{\mathbb{C}} \sum_{k=0}^{M-1}\beta_{k}|k\rangle$, 
where $\mathbb{C}=\sqrt{\sum_{k=0}^{M-1} |\beta_{k}|^{2}} $ is a normalization constant and $|k\rangle$ is the computational basis. We denote the state of the whole composite system  as $|\Phi\rangle =|\psi_s\rangle |\phi^{(t)}\rangle$ and implement the ancillary qubits controlled operations $ \sum_{k=0}^{M-1}|k\rangle \langle k|\otimes H_{k}^{g}$ on the work qubits,  and the quantum state evolves to
\begin{align}\label{entan}
	|\Phi\rangle \rightarrow \frac{1}{\mathbb{C}}\sum_{k=0}^{M-1} \beta_{k}|k\rangle H_{k}^{g} |\phi^{(t)}\rangle.
\end{align}
Then performing Hadamard gates on each qubit in the ancillary register, the state of the whole system in the subspace where ancillary register in state $|0\rangle$ evolves to
\begin{equation}
	|\Phi^{*}\rangle = \frac{1}{\mathbb{C}\sqrt{2^{m}}}|0\rangle\sum_{k=0}^{M-1} \beta_{k} H^{g}_{k} |\phi^{(t)}\rangle,
	\label{comb}
\end{equation}
which is the desired state exactly. Then we measure the auxiliary register. If $|0\rangle$ is the result, then the iteration succeeds and we obtain the desired state. This happens with a probability of $P_{s}=\parallel T|\phi^{(t)}\rangle\parallel^{2}/{\mathbb{C}^{2}M}$.
This probability can be increased by using amplitude amplification\cite{berry2015simulating} before making measurement on the auxiliary register.

\section{Results}

\subsection{Basis States and Interactions}

We adopt the harmonic oscillator basis
\begin{equation}
\begin{aligned}
	&\psi_{nlm}(\vec{r}) = R_{nl}(r)Y_{lm}(\theta, \phi) ,\\
	&R_{nl}(r) = N_{nl}e^{-\alpha^{2}r^{2}/2}(\alpha r)^{l}L^{l+1/2}_{n+l+1/2}(\alpha^{2}r^{2}) ,\\
	&N_{nl} = \sqrt{\frac{2^{l-n+2}(2l+2n+1)!!\alpha^{3}}{\pi^{1/2}n![(2l+1)!!]^{2}}}
\end{aligned}
\end{equation}
to fit the wave function, where $R_{nl}(r)$ is the radial wave function and $Y_{lm}(\theta, \phi)$ is the spherical harmonic function. $\alpha = 1.1/A^{1/6}$fm$^{-1}$, $L^{l+1/2}_{n+l+1/2}$ is the Laguerre polynomial, $n$ is the principal quantum number, $l$ and $m$ represent the orbital angular momentum and its projection on z-axis respectively. Considering also the spin part,  the eigen-states of total angular momentum are adopted
\begin{equation}
	|n,l,j,m_{j}\rangle = \sum_{m+m_{s}=m_{j}}\langle l, m; 1/2, m_{s}|j,m_{j}\rangle|\psi_{nlm}\rangle | 1/2, m_{s}\rangle,
\end{equation}
where $\langle l, m; 1/2, m_{s}|j,m_{j}\rangle$ is  Clebsch-Gordan coefficient. We define each $m_{j}$, the projection of total angular momentum single particle state, as an orbit, therefore the single particle orbit ordering is $1s^{1/2}_{-1/2}$ (orbit 1), $1s^{1/2}_{1/2}$ (orbit 2), $1p^{3/2}_{-3/2}$ (orbit 3), $1p^{3/2}_{-1/2}$ (orbit 4), $1p^{3/2}_{1/2}$ ( orbit 5),  $1p^{3/2}_{3/2}$ (orbit 6),  $1p^{1/2}_{-1/2}$ (orbit 7), $1p^{1/2}_{1/2}$ (orbit 8), $1d^{5/2}_{-5/2}$ (orbit 9) and $1d^{5/2}_{5/2}$ (orbit 10). 

The Hartree-Fock state is a good approximation to the ground state of light nuclei. In our calculation, the initial state is a superposition of Hartree-Fock state with a component of an excited state, in order to clearly show the iterative process of QCSH.

We take the interaction as given in Eq. (\ref{hamiltonian}). All the interactions are fixed except those in the Wood-saxon potential in Eq. (\ref{woodsaxon}). The average field in nuclei is the sum of the action fields of other nucleons on a single nucleon, and actually extra calculation is needed to get its exact form. Our  algorithm can serve as a subroutine to fit the parameters for nuclear models(see Supplemental for details). But to briefly illustrate the scheme of QCSH, we choose the depth of the Wood-saxon potential by fitting the binding energy calculated close to the experimental data, as shown in Table \ref{comtable}. The iterative process of QCSH is shown in Fig. \ref{first} for  four typical nuclei only here, and for the  calculations of the remaining eight nuclei, see Appendix B for details.  

\subsection{QCSH Results}
\subsubsection{Numberical Results}
We calculated the ground state energies of twelve light nuclei, $^{2}$H, $^{3}$H, $^{3}$He,$^{4}$He, $^{6}$Li, $^{7}$Li, $^{12}$C, $^{14}$N, $^{16}$O, $^{17}$O, $^{23}$Na and $^{40}$Ca. The proton and neutron single particle orbits numbers are given in Table \ref{comtable}. For instance, for $^3$H, we choose four single particle orbits for both proton and neutrons, namely, $1s^{1/2}_{-1/2}$, $1s^{1/2}_{1/2}$, $1p^{3/2}_{-3/2}$ and $1p^{3/2}_{3/2}$. For $^3$H, the Hartree-Fock state $1^{\pi}s^{1/2}_{1/2}$ $1^{\nu}s^{1/2}_{-1/2}$ $1^{\nu}s^{1/2}_{1/2}$ is a good approximation to the ground state, and in terms of qubit representation (\ref{qubitrep}), it is $|1000\rangle_{\pi}|1100\rangle_{\nu}$.  In fact, it is such a good approximation that it takes almost no iteration to converge to the final state if we start directly from the Hartree-Fock state. To demonstrate the iterative process of QCSH, we choose a superposition of Hartree-Fock state with an excited state component, $C(|1000\rangle_{\pi}|1100\rangle_{\nu}+0.01 |1010\rangle_{\pi}|1100\rangle_{\nu})$, where $C$ is the normalization constant.

\begin{figure*}
	\centering
	\includegraphics[width=0.85\linewidth]{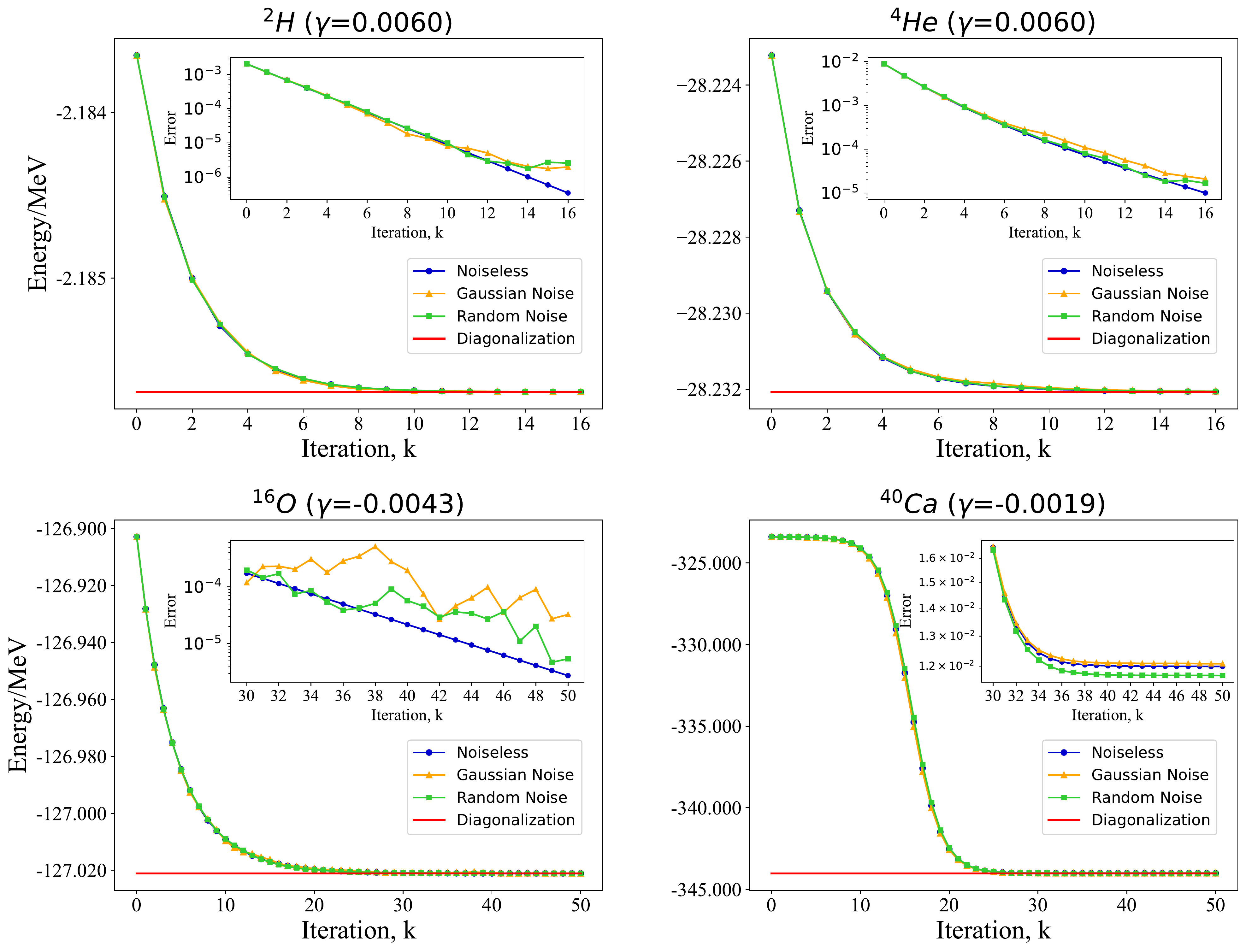}
	\caption{Iterative process to ground-state energy by QCSH for $^{2}$H, $^{4}$He, $^{16}$O and $^{40}$Ca. The initial states are superposition of Hartree-Fock state and an excited state. The exact classical diagonalization results (red line), the QCSH without noise (blue line), QCSH with a Gaussian noise(orange line), and QCSH with random noise (green line) are also shown.}
	\label{first}
\end{figure*}

We perform the quantum gradient descent iteration of Eq. (\ref{entan})  and obtain the ground state energies by numberically simulation.  The results are marked as noiseless in Fig.\ref{first}, and shown on the right-most column in Table \ref{comtable}. From the calculated results, we can see  the ground state energies of these nuclei agree with the experimental data very well. 
\begin{table}[htpb]
\centering
\begin{tabular}{c|cccc|ccc}
\hline
	$^{A}_{Z}X$ & ($N_{\pi}$, $N_{\nu}$) & \begin{tabular}[c]{@{}l@{}}Shell-Model\\ Dimensions\end{tabular} & \begin{tabular}[c]{@{}l@{}}\; $U_{0}$ \\ /MeV\end{tabular} & \begin{tabular}[c]{@{}l@{}} QCSH\\ Qubits\end{tabular} & \multicolumn{2}{c}{\begin{tabular}{cc}
\multicolumn{2}{c}{\; Binding Energy/MeV \; }\\
\quad \, \; Exp. & \; \, QCSH \\
\end{tabular}} \\ \hline
	$^{2}$H & (4, 4) & $2^{8}\times 2^{8}$ & -48.0 & 15 & \quad \, -2.22 & -2.19 \\
	$^{3}$H & (4, 4) & $2^{8}\times 2^{8}$ & -45.4 & 15 & \quad \, -8.48 & -8.67 \\
	$^{3}$He & (4, 4) & $2^{8}\times 2^{8}$ & -45.4 & 15 & \quad \, -7.72 & -7.59  \\
	$^{4}$He & (4, 4) & $2^{8}\times 2^{8}$ & -42.9 & 15 & \quad \, -28.30 & -28.23  \\
	$^{6}$Li & (6, 6) & $2^{12}\times 2^{12}$ & -40.6 & 21 & \quad \, -31.99 & -31.45 \\
	$^{7}$Li & (6, 6) & $2^{12}\times 2^{12}$ & -40.6 &  21 & \quad \, -39.24 & -39.46 \\
	$^{12}$C & (8, 8) & $2^{16}\times 2^{16}$ & -38.9 & 27 & \quad \, -92.16 & -95.03 \\
	$^{14}$N & (8, 8) & $2^{16}\times 2^{16}$ & -38.9 & 27 & \quad \, -104.66 & -108.36 \\
	$^{16}$O & (10, 10) & $2^{20}\times 2^{20}$ & -38.9 & 32 & \quad \, -127.61 & -127.02 \\
	$^{17}$O & (12, 12) & $2^{24}\times 2^{24}$ & -38.4 & 37 & \quad \, -131.76 & -130.83 \\
	$^{23}$Na & (14, 14) & $2^{28}\times 2^{28}$ & -37.5 & 42 & \quad \, -186.56 & -185.35 \\
	$^{40}$Ca & (22, 22) & $2^{44}\times 2^{44}$ & -37.6 & 61 & \quad \, -342.05 & -344.02 \\ \hline
\end{tabular}
\caption{The number of single particle orbits for protons and neutrons($N_{\pi}$, $N_{\nu}$), the dimension of matrix to be diagonalized  in classical shell-model calculation ( Shell-model Dimensions), the number of qubits used in QCSH (QCSH Qubits), Experimental Binding Energy (Exp.), and QCSH calculated binding energy.}
\label{comtable}
\end{table}

To simulate the noises in practical quantum computing device, we added a noise term $\sum_{k=1}^{A}\delta \alpha_{k} Z_{k}$ into the Hamiltonian of the system, and a perturbation $|\delta \psi\rangle$ to the quantum state $|\psi^{(t)}\rangle$. Namely, the quantum state is truly $|\tilde{\psi}^{(t)}\rangle=(|\psi^{(t)}\rangle + |\delta \psi\rangle)/|||\psi^{(t)}\rangle + |\delta \psi\rangle||$. $\delta \alpha_k$ and $|\delta \psi\rangle$ are chosen both as Gaussian distribution, $P(x)=e^{-(x-\mu)/2\sigma^2}/\sqrt{2\pi\sigma}$ with $\mu = 0$,  $\sigma=0.1/3$ , which are marked as Gaussian noise in  Fig. \ref{first}; or both as  uniform distribution with an amplitude of 0.02, which are marked as random noise in Fig. \ref{first} respectively. In this case, the iteration operator is $\tilde{T}=I-2\gamma(H+\sum_{k=1}^{A}\delta \alpha_{k} Z_{k})$.  Because  the measurement at the final step may obtain wavefunctions with different number of particles, a projection is performed  so as to conserve the proton and neutron particle numbers.

The QCSH finds the eigenvalue through iteration quickly. As shown in Fig. \ref{first}, it takes  less than 20 iteration to reach the ground state. We compared the results of QCSH with the exact diagonalization of the noiseless Hamiltonian in a classical computer. The classical exact results are given as Diagonalization in Fig. \ref{first}.   The results of our quantum algorithm without noise are consistent with classic diagonalization. For cases with  a given strength of noise, the results of QCSH  converge to the exact energy of diagonalization within an error of about 1 KeV, which is within the accuracy for nuclear structure physics. In fact,  QCSH converges more quickly if the initial states are taken as Hartree-Fock state \cite{ring2004nuclear}. This noise analysis demonstrated that QCSH is robust against noises, and is very promising to be realized in NISQ quantum computers in the near future.
\subsubsection{Experimental Results}
\begin{figure}
	\centering
	\includegraphics[width=0.9\linewidth]{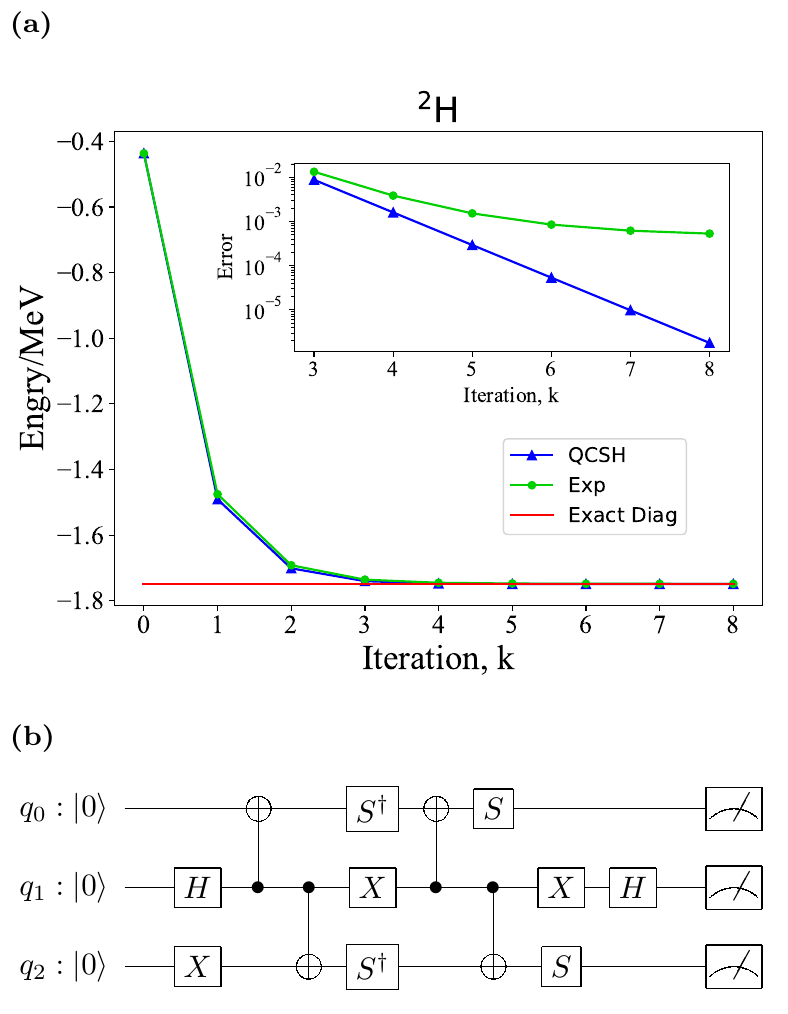}
	\caption{(a) The iteration process of calculating deuteron binding energy. The initial states are Hartree-Fock state  $| n\rangle=|10\rangle $. (b)Quantum circuit for calculating $\langle m | H_{xy} | n\rangle$. }
	\label{experiment}
\end{figure}

We carried out the experiment of simulating deuteron on a superconducting computer.  Our results are based on cloud access to the ScQ .Cloud\cite{SCQ}, which consist of 10 superconducting qubits. The Hamiltonian of deuteron calculated by QCSH contains  8 qubits.  Limited by the existing realities of cloud quantum computing, we have to adjust
the employed Hamiltonian of deuteron \cite{dumitrescu2018cloud}as 
\begin{eqnarray}\label{H2}
H_2 = 5.906709 I +0.218291Z_0 -6.125 Z_1 \\\nonumber
-2.143304 (X_0 X_1 + Y_0Y_1),
\end{eqnarray}
that can be claculated with 3 superconducting qubits. 
We choose $| n\rangle=|10\rangle $ as initial state and $\gamma=0.02 $.  We have $(I-2 \gamma H_2)| n\rangle =(I-2 \gamma H_0)| n\rangle+(I-2 \gamma \sum_m \langle m | H_{xy} | n\rangle) | m\rangle$, where $H_0=5.906709 I +0.218291Z_0 -6.125 Z_1$ and $H_{xy} =-2.143304 \left(X_0 X_1 + Y_0Y_1\right)$. The value $\langle m | H_{xy} | n\rangle$ can be calculated by the following quantum circuit in Fig. \ref{experiment}(b). To militate  the effect of noise, we perform 20000 measurements for each calculation, and the results are also shown in  Fig. \ref{experiment}(b).  The binding energy $E({H_2})= -1.748\pm 0.02$~MeV, agree with the exact energy of $-1.749$~MeV within uncertainties. 

\subsection{Computational Complexity of QCSH}

We first analyze the qubit resource complexity. In a nucleus, if the proton occupies $N_{\pi}$  single particle states and neutron occupies $N_{\nu}$  single particle states, the number of qubits for representing quantum state of the system is $N_{\pi} + N_{\nu}$. For simplicity, we set $N_{\pi}=N_{\nu}=\mathcal{N}$ to characterize the system size.  The number of Pauli product terms from one-body interaction is less than $ O(\mathcal{N}^2)$, and the number of Pauli product terms from two-body integrals is no more than $ O(\mathcal{N}^4)$. The number of ancillary qubits is
\begin{equation}
	m = \lceil \log_{2} M \rceil \leq \lceil \log_{2}(12\mathcal{N}^{2}(2\mathcal{N}-1)^{2}+4\mathcal{N}(2\mathcal{N}-1)) \rceil.
\end{equation}
Thus total number of required qubits is $O(2\mathcal{N}+2m-1)$.

As for gate complexity, we need to decompose the many-qubit control gates $ \sum_{k=0}^{M-1}|k\rangle \langle k|\otimes H_{k}^{g}$  into basic qubit gates. If we restrict to only one-body and two-body interactions,  a $m$-qubit controlled Pauli gate $C^{m}(U)$ can be decomposed of $M_{c}=32(m-1)+4$ basic gates \cite{nielsen2000quantum}. For each iteration, the complexity is $ O(\mathcal{N}^{4}\log_{2}\mathcal{N})$.
By contrast, diagonalizing a $2^{\mathcal{N}}$-dimension symmetric matrix on a classic computer using the symmetric Gaussian elimination takes $O(2^{3\mathcal{N}})$ steps. Therefore, in the circumstances where iteration steps $k \in O(\text{poly} N)$, QCSH achieves expoential speedup compared with classical Shell model.

\section{Discussion}
QCSH  is an efficient quantum algorithm to calculate the nuclear structure. The gate complexity and qubit resources  are both polynomial to the number of selected single particle states. With the wavefunctions at hands, one can calculate the transition rates, reaction cross section and so on directly. 

In QCSH, we usually  obtain the target state with a certain probability, as is true for a LCU algorithm. The convergence rate depends on the appropriate choice of the step size factor, $\gamma$, and the initial state. Selecting Hartree-Fock state as the initial state can  significantly accelerate the convergence,  and mitigate the error. Another effective way is to apply amplitude amplification to increase the probability of success,and it will be useful when the depth of quantum circuits in hardware becomes larger.

We have demonstrated QCSH by applying it to twelve light nuclei. The results are completely consistent with  those  from exact diagonalization on a classical computer.  Although near-term quantum computer hardware still has errors, and can work only for shallow gate operations, it can already perform some meaningful tasks. Most importantly, progress in hardware is  rapid and accelerated in recent years. Application of QCSH will be smoothly transformed into  future large-scale fault-tolerant quantum computers due to its full quantum computing nature. 

\textbf{ Acknowledgements}\\
We gratefully acknowledges support from the National Natural Science Foundation of China under Grants No. 11974205, and No. 12005015. The National Key Research and  Development Program of China (2017YFA0303700); The Key Research and  Development Program of Guangdong province (2018B030325002); Beijing Advanced Innovation Center for Future Chip (ICFC).
\bibliography{qcsh}
\bibliographystyle{unsrt}
\pagebreak
\section{Appendix A:Details about the principle of QCSH }\label{AB}

Gradient descent is an iterative algorithm to find a local minimum of a target function. In details, Gradient embodies the local properties of the target function, and it represents the fastest growing direction, so the target function decreases most fastly against the gradient direction.

As for target function $f(\vec{\theta})$, the iteration formula is
\begin{equation}
\theta^{(t+1)} = \theta^{(t)} - \eta \nabla_{\theta} \left. f(\theta)\right|_{\theta=\theta^{(t)}}
\end{equation}

where $\eta$ is the iteration step or learning rate. Because gradient could only embody local properties, $\eta$ should be small compared to the scale of parameter space we care, or the algorithm might be hard to converge. There is a trade-off between convergence and iterative efficiency.

The Hamiltonian of a quantum system is usually Hermite, and their eigenvalues are all real numbers. Consider a $d$-dimension system, assume the eigenstates and eigenvalues are $\{(|u_{k}\rangle, \lambda_{k})\}_{k=0}^{d-1}$, and eigenvalues have been sorted as $\lambda_{0}\leq\lambda_{1}\leq...\leq\lambda_{d-1}$. Any initial state can be decomposed as linear superposition of eigenstates of the system
\begin{equation}
	|\phi^{(0)}\rangle = \sum_{k=0}^{d-1}c_{k}|u_{k}\rangle
\end{equation}
consective implement the  generate iteration operator $T = I-2\gamma H$ on the quantum state, we have
\begin{equation}
	|\phi^{(t)}\rangle = \frac{1}{\mathbb{D}}\sum_{k=0}^{d-1}c_{k}(1-2\gamma\lambda_{k})^{t}|u_{k}\rangle,
\end{equation}
where  $\mathbb{D}=\sqrt{\sum_{k=0}^{d-1}|c_{k}(1-2\gamma\lambda_{k})^{t}|^2}$ is a normalization constant. It is not hard to find that $|\phi^{(t)}\rangle$ approach to $|u_{k^{*}}\rangle$ which satisfies
\begin{equation}
	k^{*}=\mathop{argmax}\limits_{c_{k}\neq 0} |1-2\gamma \lambda_{k}|.
\end{equation}

If we want to obtain the ground state of the system, we will show that this can be achieved by reasonably selecting the value of $\gamma$. Firstly we estimate an upper bound of the smallest and largest eigenvalue 
\begin{equation}
	\lambda_{0} < q, \; \lambda_{d-1} < Q,
\end{equation}
then we have $|\frac{q+Q}{2}-\lambda_{0}|\geq |\frac{q+Q}{2}-\lambda_{k}|$ for any $k \in \{0, 1, ..., d-1\}$. To ensure $|u_{0}\rangle$ is the target state of iteration, $|1-2\gamma\lambda_{0}|$ must be the biggest among $\{|1-2\gamma \lambda_{k}|\}_{k=0}^{d-1}$ or in other words $\frac{1}{2\gamma} > \frac{q+Q}{2}$ should be satisfied, which derivates
\begin{equation}
\gamma \in \left\{
	\begin{aligned}
		& \quad (0, \frac{1}{q+Q}), & q+Q>0,\\
		& \quad (0, +\infty),  & q+Q = 0 ,\\
		& \quad (-\infty, \frac{1}{q+Q}) \cup (0, +\infty), & q+Q<0.
	\end{aligned}
\right.
\end{equation}

Because $\lambda_{0} \leq \lambda_{d-1}$, we can always choose $q=Q$. Estimating an upper bound of $\lambda_{0}$ is dispensible. While for the sake of efficiency, the closer $q$ and $Q$ are to the exact bound, the higher the iteration efficiency.

When the ground state is degenerate, this algorithm could not distinguish the degenerate states, but it is still reliable to calculate the ground state engry.

\begin{figure}
	\centering
	\includegraphics[width=0.9\linewidth]{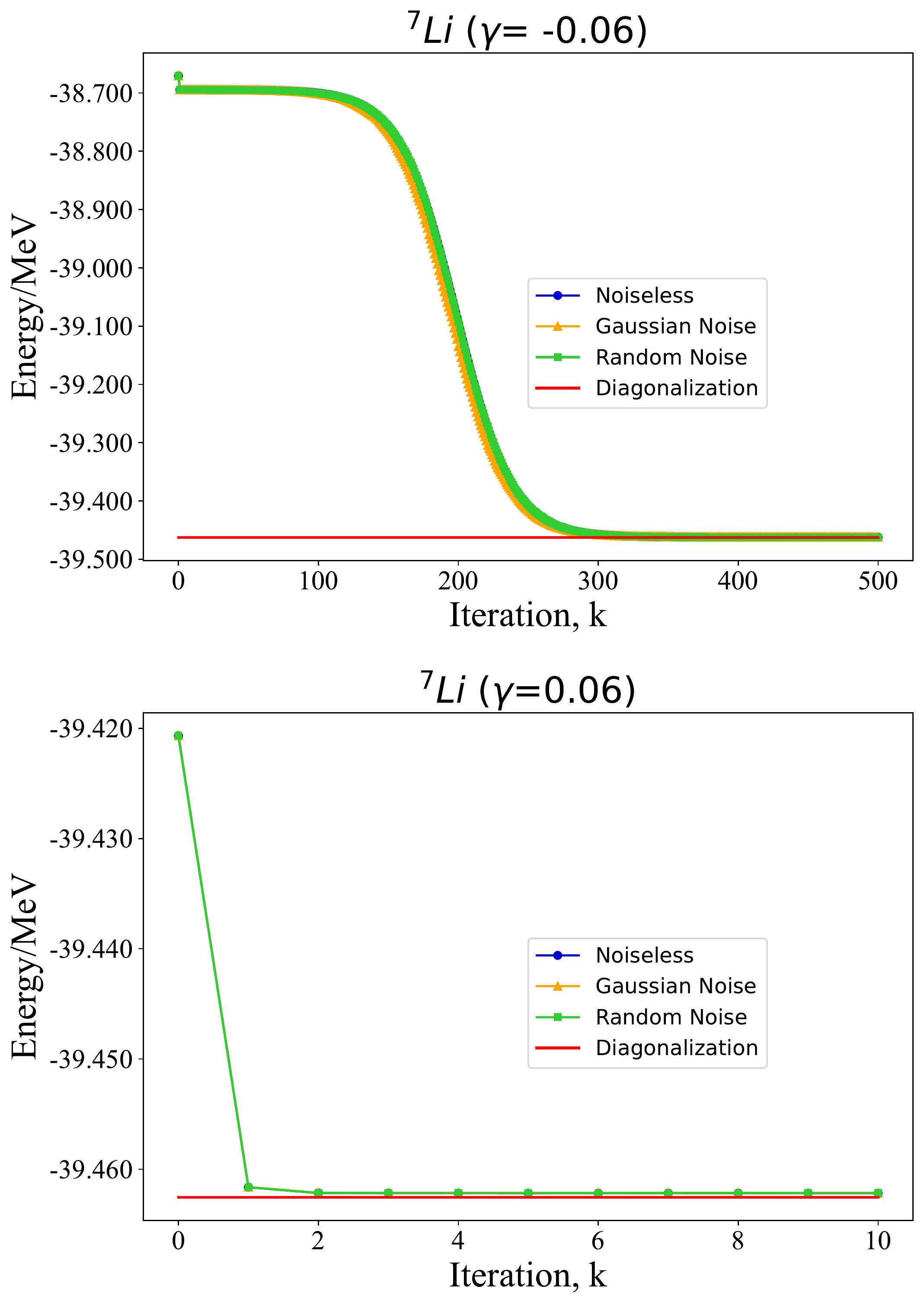}
	\caption{The comparasion of iteration on different initial states for $^{7}$Li. The initial states of QCSH are selected as $|\phi^{(0)}\rangle \propto (|111000\rangle _{\pi}|111100\rangle _{\nu} + 0.01 |110001\rangle _{\pi}|110110\rangle _{\nu})$ and $|\phi^{(0)}\rangle \propto (|110001\rangle _{\pi}|110110\rangle _{\nu} - |110001\rangle _{\pi}|111001\rangle _{\nu})$ respectively.  Even though the latter evolves with a larger $\gamma$, it converges to the ground state energy faster.}
	\label{iterationcompare}
\end{figure}

Another problem is about the initial state. Even if we choose a reasonable $\gamma$, the condition $c_{0}=\langle u_{0}|\phi^{(0)}\rangle \neq 0$ has to be satisfied, or we could not obtain the ground state. Fortunately, the Hartree-Fock state $|HF\rangle$ is already a good approximation to ground state, which usually satisfies $\langle u_{0}|HF\rangle \neq 0$. But for some nuclei with mixed configurations, $|HF\rangle$ might not be a good choice, because the efficiency of iteration will be extremely low as shown in Fig. \ref{iterationcompare}. In this case, other configurations should be added with a larger weight.

\section{Appendix B: Numberical  calculations of the remaining eight nuclei}
\begin{figure}[h]
	\centering
	\includegraphics[width=0.9\linewidth]{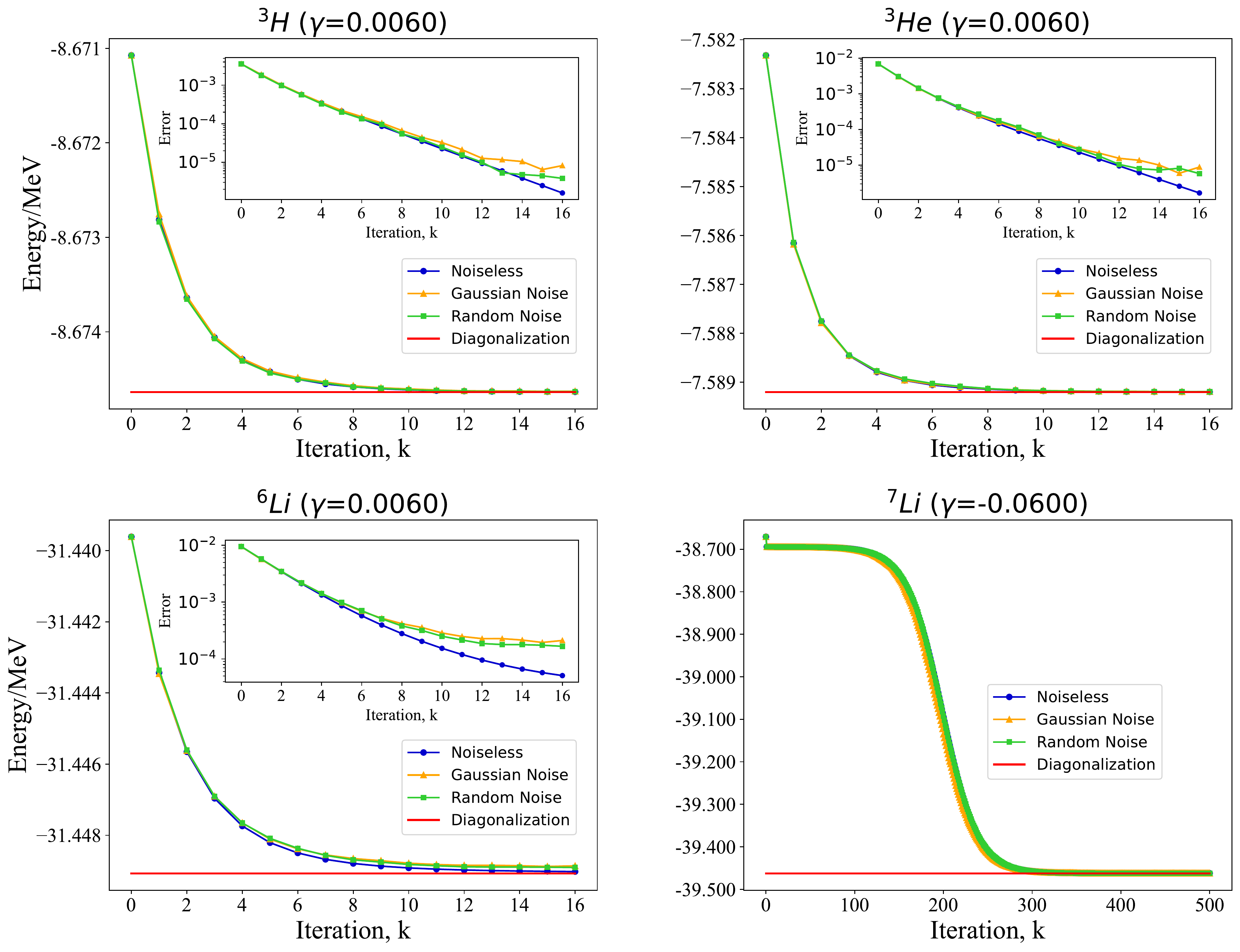}
	\caption{Iterative process to ground-state energy by QCSH for $^{3}$H, $^{3}$He, $^{6}$Li and $^{7}$Li.}
	\label{second}
\end{figure}

\begin{figure}[h]
	\centering
	\includegraphics[width=0.9\linewidth]{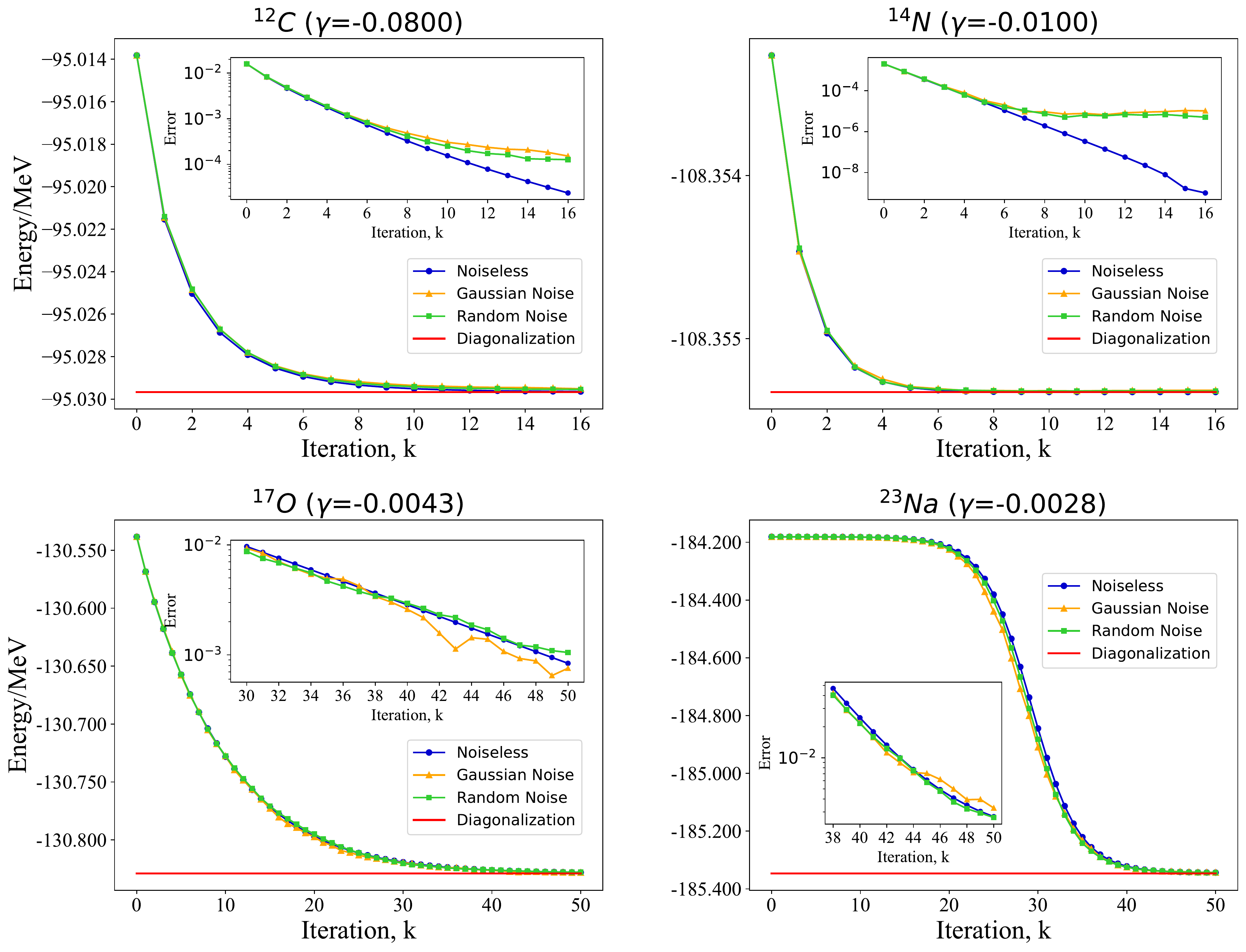}
	\caption{Iterative process to ground-state energy by QCSH for $^{12}$C, $^{14}$N, $^{17}$O and $^{23}$Na.}
	\label{third}
\end{figure}

Here we show  the iterative process of QCSH for eight typical nuclei  in Fig. \ref{second} and Fig. \ref{third}. The initial states are choosen as superposition of Hartree-Fock state and an excited state. The exact classical diagonalization results (red line), the QCSH without noise (blue line), QCSH with a Gaussian noise(orange line), and QCSH with random noise (green line) are also shown.

\section{Appendix C: Fitting Parameters for Nuclear Models}
	\; \; Our  algorithm can serve as a subroutine to fit the parameters for nuclear models. When the form of a nuclear structure model is obtained through theoretical analysis, parameters of this model should be well selected to conform to the experimental results. As the simulation of nuclei on classic computers is intractable, the process of adjusting parameters is in high complexity. However, employing quantum computation to adjust the parameters can raise the efficiency of this process.

	In the maintext, we directly use the average field depth $U_{0}$ with properly fixed  parameters to examine the feasibility of our algorithm by calculating $^{3}$H, $^{3}$He, $^{6}$Li, $^{12}$C, $^{14}$N. Here we will show an approximate method to obtain the parameters above based on our algorithm. After that, we use the parameters fitted to predict the ground state energy of $^{7}$Li and $^{16}$O. Then a comparison between theory and experiments is introduced, the results inspect the applicability of the nuclear shell model.

	To fit the parameters in $U_{0} = (u+a\cdot(N-Z)/A+b \cdot Z + c \cdot N + d / (N + Z)) \;$ MeV, for each $^{A}_{Z}$X in \{$^{3}$H, $^{3}$He, $^{6}$Li, $^{12}$C, $^{14}$N\},  we variate the value of $U_{0}$, and calculate the ground state energy corresponding to a specific $U_{0}$ using FQE. Denoting the value of $U_{0}$ as $U^{*}$,  when  the theoretical result obtained by FQE  is equal to the ground state energy obtained in experiments. We obtain the $U^{*}$ for every nucleon and use them to fit the parameters by linear regression. Meanwhile, we record the upper bound and lower bound of $U_{0}$ with the allowable error  $0.3$MeV/nucleon, in other words, when $U_{0}$ is within this value range, difference between prediction of the shell model and experiments is less than $0.3$MeV average to each nucleus. Results are shown in Table \ref{fitdata}, which indicate the appropriate average field depth roughly increases with the number of nucleons.

	\begin{table}[htpb]
	\centering
	\begin{tabular}{c|ccc|c}
	\hline
		$^{A}_{Z}X$ & $E_{0}$/MeV & ($E_{0}/A$)/MeV & $U^{*}$/MeV & Allowable Depth/MeV \\ \hline
		$^{3}H$ & -8.48 & -2.83 & -45.30 & -45.78 $\sim$ -44.83 \\
		$^{3}He$ & -7.72 & -2.57 & -45.47 & -45.94 $\sim$ -44.99 \\
		$^{6}Li$ & -31.99 & -5.33 & -40.71 & -41.10 $\sim$ -40.32 \\
		$^{12}C$ & -92.16 & -7.68 & -38.61 & -38.97 $\sim$ -38.26 \\
		$^{14}N$ & -104.66 & -7.48 & -38.57 & -38.95 $\sim$ -38.22 \\ \hline
	\end{tabular}
	\caption{Vaule range of $U_{0}$ to average field potential for nucleus within a given error $0.3$MeV/nucleon. $E_{0}$ is the ground state energy in physical experiments. $A$ is the nucleon number for each nucleus. When $U_{0}$ is within allowable depth, the difference between theoretical result and experimental data of $E_{0}$/A is less than $0.3$MeV. }
	\label{fitdata}
	\end{table}
	
	In order to obtain the parameters efficiently, we set a loss function for linear regression, and optimize the parameters by mimizing this loss function
	\begin{equation}
		\text{Loss} = \sum_{k}[(u+a\cdot(N_{k}-Z_{k})/A_{k}+b \cdot Z_{k} + c \cdot N_{k} + d / A_{k}) - U_{k}^{*}]^{2}.
		\label{fitformula}
	\end{equation}
	where $k \in \{$ $^{3}$ H, $^{3}$ He, $^{6}$Li, $^{12}$C, $^{14}$N $\}$.

\begin{figure}[h]
		\centering
		\includegraphics[width=0.9\linewidth]{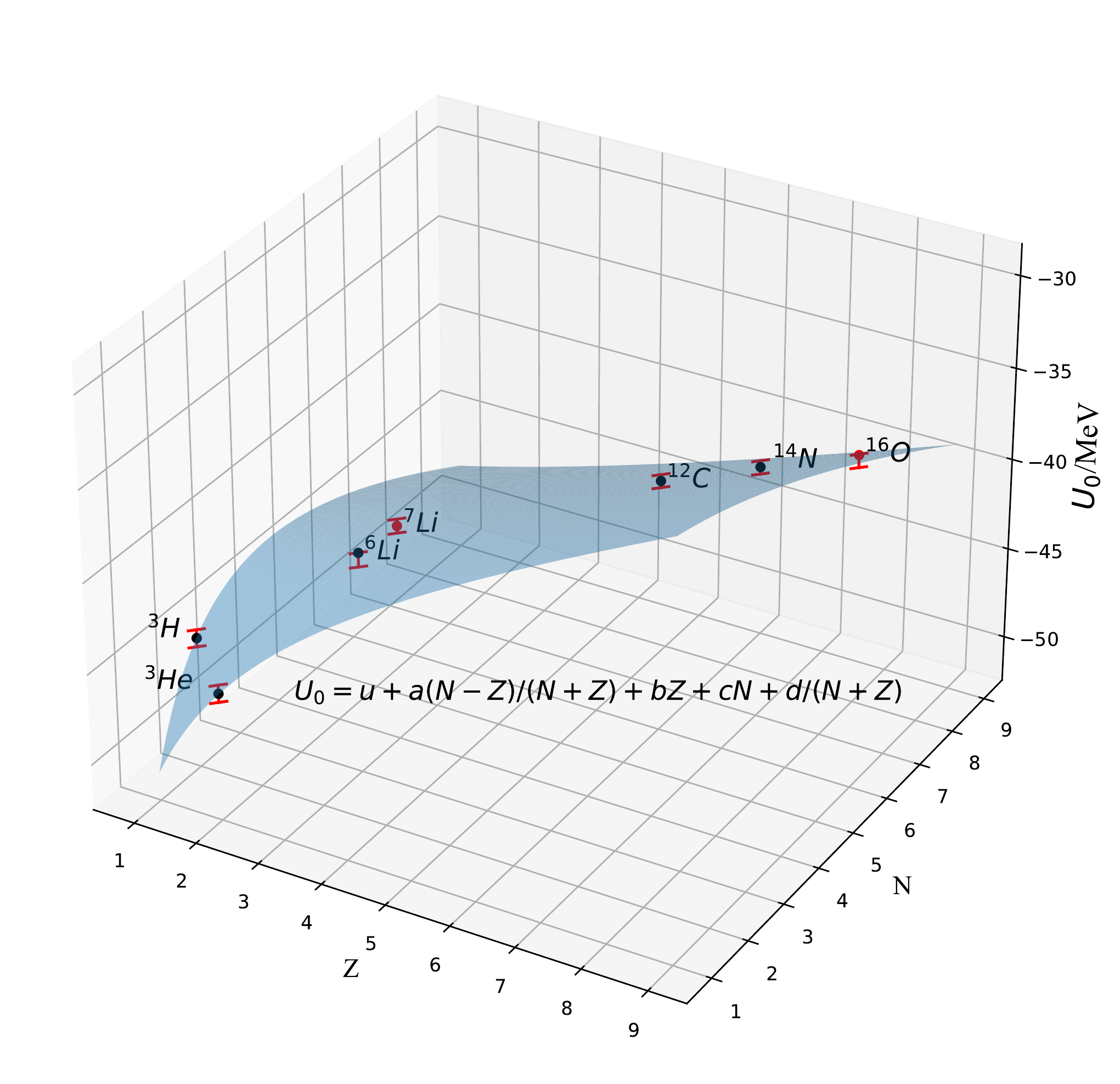}
		\caption{Fitting for average field depth $U_{0}$. The black points and red points are calculated by the fitting formula, which are exactly on the sphere of $U_{0} = u+a\cdot(N-Z)/A+b \cdot Z + c \cdot N + d / A = (-33.65 + 5.175 \cdot (N - Z) / A + 1.46 \cdot Z -1.82 \cdot N - 33.57 / A) \;$ MeV. The red error bar represents the value range of $U_{0}$ compared to experimental data in physics for each nucleus.}
		\label{fitgraph}
	\end{figure}

	After fitting the parameters of the average field, we use the formula fitted to caculate $U_{0}$ for  $^{7}$Li and $^{16}$O, and compare the results of average filed approximation and experiments in Table \ref{comtable} and Figure \ref{fitgraph}.
	The simulating results show that predictions of ground state energy for $^{7}$Li and $^{16}$O are also within the error range compared to experiments in nuclear physics, as shown in Figure \ref{fitgraph} with red points. With the parameters fitted using experimental data of few nuclei, the nuclear shell model can make a good prediction of more complex nucleus. Our quantum algorithm, as a subroutine, provides an economic method to fit the parameters for nuclear models.

\begin{figure}[h]
		\centering
		\includegraphics[width=0.9\linewidth]{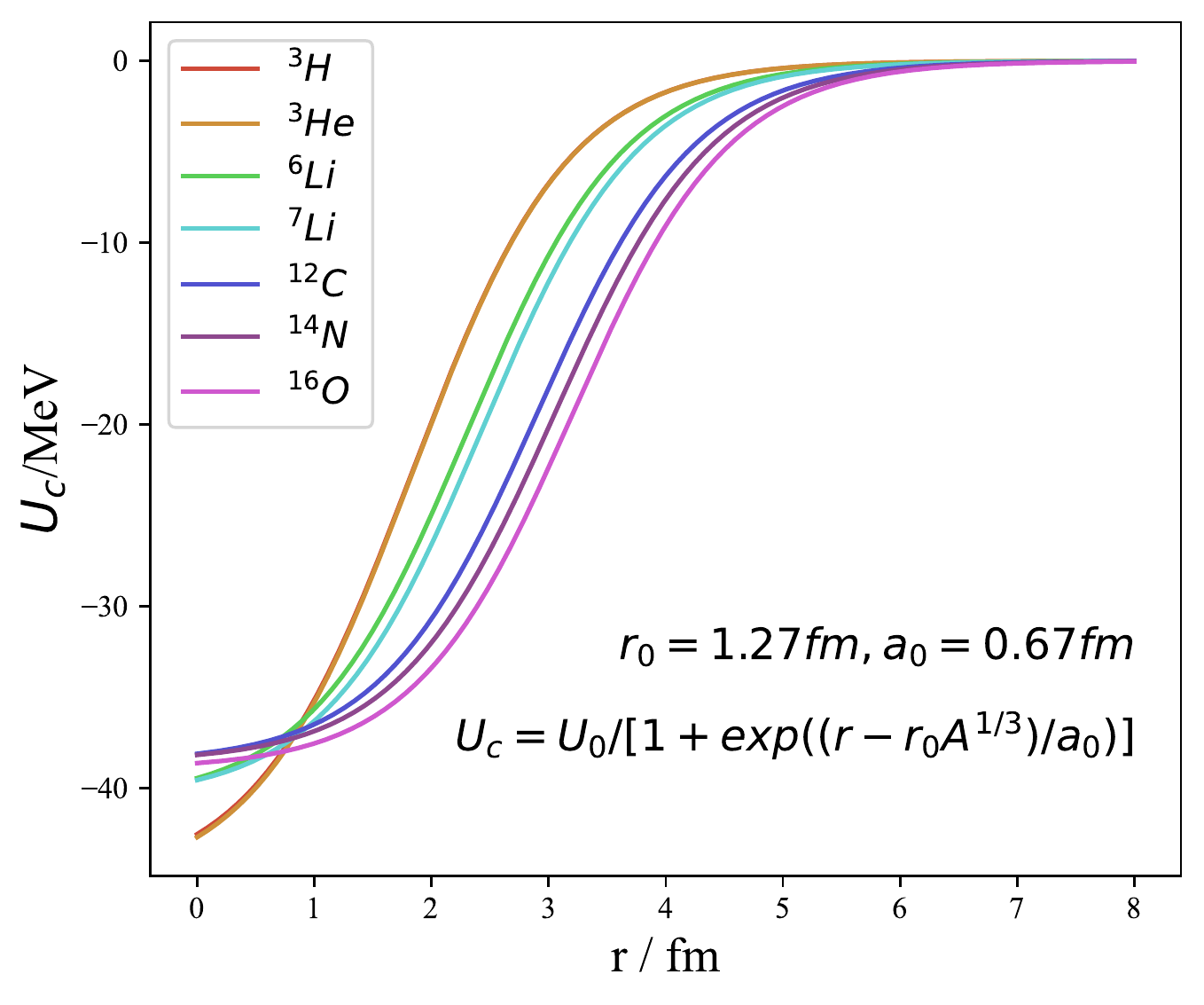}
		\caption{Woods-Saxon potential in different nucleus. $U_{0}$ is set as $U^{*}$ for every nucleus.}
		\label{meanfield}
	\end{figure}

	The details for Woods-Saxon potential in different nucleus are shown in Fig. \ref{meanfield}.

	\section{Appendix D: Computational complexity}
\begin{figure}[h]
			\centering
			\includegraphics[width=0.9\linewidth]{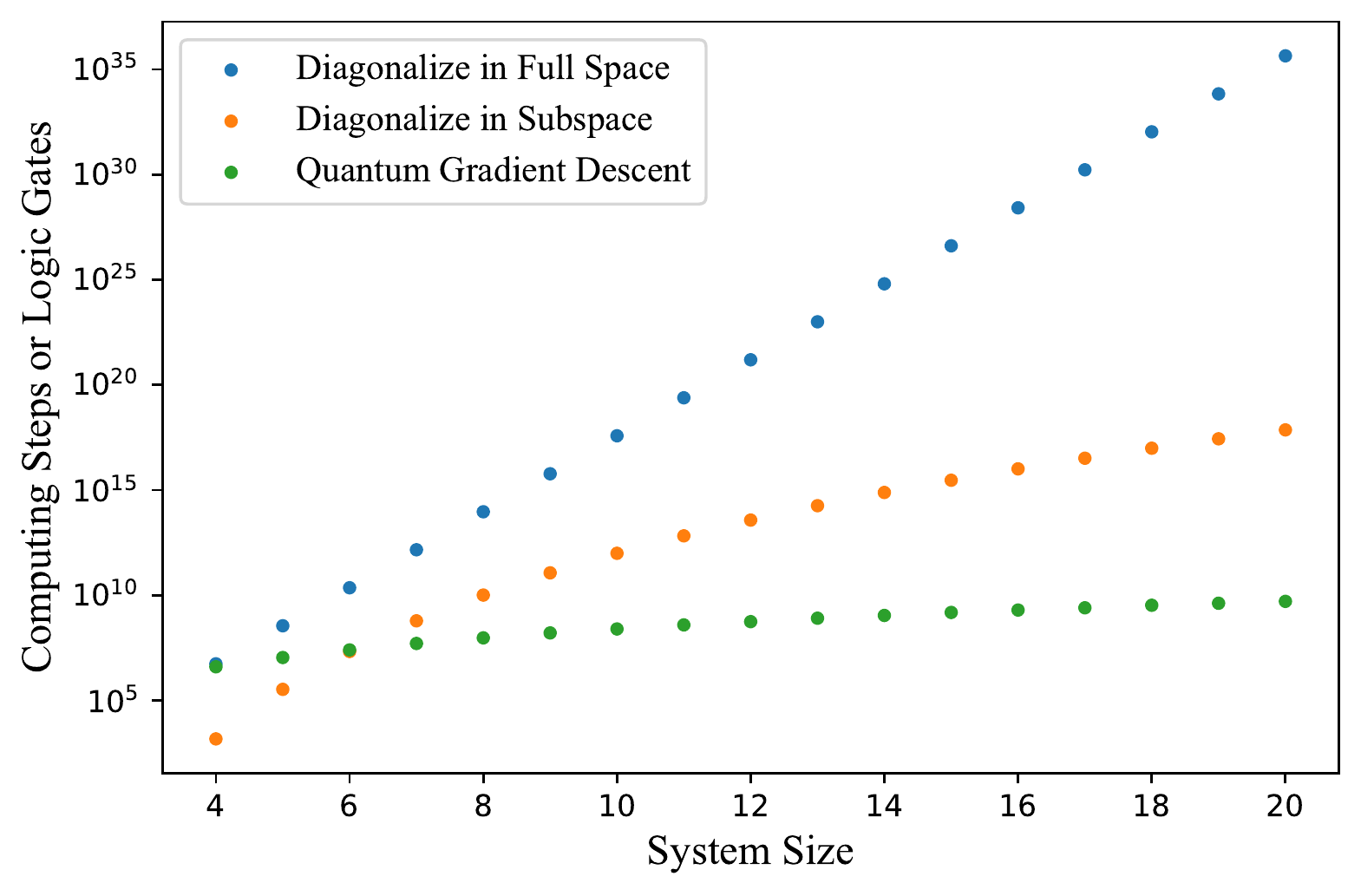}
			\caption{Comparison of gate complexity between classic diagonalization of Hamiltonian and FQE.}
			\label{gate}
\end{figure}

	Here we list the qubits resources and gate complexity for simulating the nucleus in Fig. \ref{gate}.  Here the subspace is composed of the basis that satisfy the conservation of proton number and neutron number respectively, and the other basis can be ignored for a system with a certain number of particles. The dimensions of subspace are usually small than that of full space among the many-body systems.

\end{document}